\documentclass[twocolumn,showpacs,preprintnumbers,amsmath,amssymb,times,pre]{revtex4}

\usepackage{graphicx}
\usepackage{dcolumn}
\usepackage{bm}
\usepackage{amsmath}

\begin{document}

\title{Universal role of migration in the evolution of cooperation}

\author{Han-Xin Yang$^{1}$}\email{hxyang@mail.ustc.edu.cn}

\author{Wen-Xu Wang$^{2}$}\email{wenxuw@gmail.com}
\author{Bing-Hong Wang$^{1,3}$}\email{bhwang@ustc.edu.cn}

\affiliation{$^{1}$Department of Modern Physics, University of
Science and Technology of China, Hefei 230026,
China\\$^{2}$Department of Electronic Engineering, Arizona State
University, Tempe, Arizona 85287-5706, USA\\$^{3}$The Research
Center for Complex System Science, University of Shanghai for
Science and Technology and Shanghai Academy of System Science,
Shanghai, 200093 China }

\begin{abstract}

We study the role of unbiased migration in cooperation in the
framework of spatial evolutionary game on a variety of spatial
structures, involving regular lattice, continuous plane and complex
networks. A striking finding is that migration plays a universal
role in cooperation, regardless of the spatial structures. For high
degree of migration, cooperators cannot survive due to the failure
of forming cooperator clusters to resist attacks of defectors. While
for low degree of migration, cooperation is considerably enhanced
compared to statically spatial game, which is due to the
strengthening of the boundary of cooperator clusters by the
occasionally accumulation of cooperators along the boundary. The
cooperator cluster thus becomes more robust than that in static game
and defectors nearby the boundary can be assimilated by cooperators,
so the cooperator cluster expands, which facilitates cooperation.
The general role of migration will be substantiated by sufficient
simulations associated with heuristic explanations.

\end{abstract}

\date{\today}


\maketitle

Cooperation is fundamental to biological and social systems.
Understanding factors that facilitate and hamper cooperation is a
significant issue. In the framework of evolutionary games, a number
of mechanisms in favor of cooperation have been found, such as
costly punishment~\cite{punishment1,punishment2},
reputation~\cite{reputation1,reputation2} and social diversity
~\cite{diversity1,diversity2,diversity3}. Quite recently, the role
of migration in cooperative behavior has drawn growing
interests~\cite{migration1,migration2,migration3,success-driven,migration4,migration5},
because of the fact that migration is a common feature in nature and
society. For example, millions of animals migrate in the savannas of
Africa every year, and thousands of people travel among different
countries every day. In this regard, Vainstein $et$
$al.$~\cite{migration3} considered a scenario that individuals can
move to neighboring sites on a two-dimensional lattice randomly with
some probability. In particular, it is found that such movement can
maintain and even enhance cooperation compared to the absence of
migration. More recently, Meloni $et$ $al.$ studied evolutionary
games composed of mobile players on a continuous
plane~\cite{migration5}. Their results showed that cooperation can
survive provided that both the temptation to defect and the velocity
at which individuals move are not too high. Beyond random migration,
Helbing and Yu proposed a success-driven migration strategy which is
spurred by the pursuit of profit as a nature of
individuals~\cite{success-driven}. Specifically, individuals tend to
move to neighboring site with the highest estimated payoffs.
Interestingly, such migration results in the outbreak of cooperation
in a noisy environment.

Although it has been demonstrated that migration can promote
cooperation in evolutionary games on some regular spatial
structures~\cite{migration3, migration4}, the role of migration on
other kinds of structures, for instance, complex networks, is
unknown yet. A natural concern is then whether migration plays
some general role in cooperation, regardless of underlying
structures, or there exists dependence of the role of migration on
structures? To address this issue, in this paper we incorporate
random migration in evolutionary games on a variety of spatial
structures, involving continuous space, regular structure, and
typical complex topologies. Strikingly, we find that the role of
random migration in promoting cooperation is universal, regardless
of different structures. This is somewhat counterintuitive in the
sense that mobility of individuals may weaken the stability of
cooperation clusters which are key for the survival of
cooperators. However, we will substantiate the positive effect of
random migration on cooperation by intensive simulations and
provide convinced explanations for the underlying mechanisms.

To be concrete, we use the Prisoner's Dilemma \cite{prisoner's
dilemma} to carry out our researches. In principle, the Prisoner's
Dilemma is a game played by two players, each of whom chooses one of
two strategies, cooperation or defection. They both receive payoff
$R$ upon mutual cooperation and $P$ upon mutual defection. If one
defects while the other cooperates, cooperator receives $S$ while
defector gets $T$. The ranking of the four payoff values is: $T>R>
P>S$. Thus in a single round of the Prisoner's Dilemma it is best to
defect regardless of the opponent's decision. The Prisoner's Dilemma
has attracted much attention in theoretical and experimental studies
of cooperative behavior. Following common practice \cite{spatial 1},
we set $T=b$ $(1<b<2)$, $R=1$, and $P=S=0$, where $b$ represents the
temptation to defect.

\begin{figure*}
\begin{center}
\scalebox{0.85}[0.85]{\includegraphics{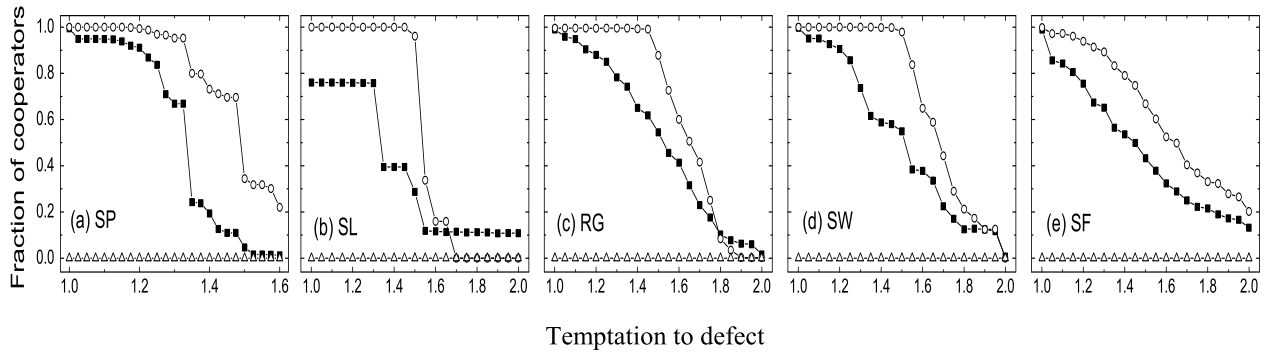}}
\caption{Fraction of cooperators as a function of the temptation to
defect $b$. (a) Individuals are located on a continuous square plane
(SP) of linear size $L=20$ with periodic boundary conditions.
Interaction radius $q=1$. Filled squares, open circles and open
triangles show results for $v=0$, $v=0.04$ and $v=1$ respectively.
Lines are guided for eyes. (b)-(e) Individuals are located on square
lattices (SL), random graphs (RG), small-world networks (SW) and
scale-free networks (SF). Average connectivity $z=4$ for SL and SF,
$z=6$ for RG and SW. Filled squares, open circles and open triangles
show results for $p=0$, $p=0.001$ and $p=1$ respectively. The
population size is 1024. The equilibrium fraction of cooperators
results from averaging over $10^{5}$ time steps after a transient
period of $10^{5}$ time steps. Each data point depicted corresponds
to an average over 1,000 simulations; that is, 100 runs for 10
different realizations of the same class of graph.} \label{fig:the
fraction of cooperators}
\end{center}
\end{figure*}

To explore the role of migration, we resort to the spatial game in
which individuals are placed on some spatial structures. Since the
combination of spatial structures into evolutionary games by Nowak
and May~\cite{spatial 1}, there has been much interest in revealing
the influence of population structures on cooperation, ranging from
regular lattices to complex networks \cite{spatial 2,spatial
3,spatial 4,spatial 5,spatial 6,spatial 7,spatial 8,spatial
9,spatial 10,spatial 10.1,spatial 11,spatial 12,spatial 13,spatial
14,spatial 15}. In the spatial games, interactions among individuals
are restricted within immediate neighbors and usually neighbors of
an arbitrary individual keep fixed. While in the presence of
migration, neighboring individuals can be changed by encountering
different partners as time goes on. In the seminal works of
Vainstein $et$ $al.$~\cite{migration3} and Meloni $et$
$al.$~\cite{migration5}, random migration has been considered in
spatial games on a two-dimensional lattice and on a continuous
plane, respectively. Inspired by these original researches, we
extend migration on regular structure to complex networks and
uncover the general role of migration in promoting and hampering
cooperation.

\begin{figure}
\begin{center}
 \scalebox{0.84}[0.84]{\includegraphics{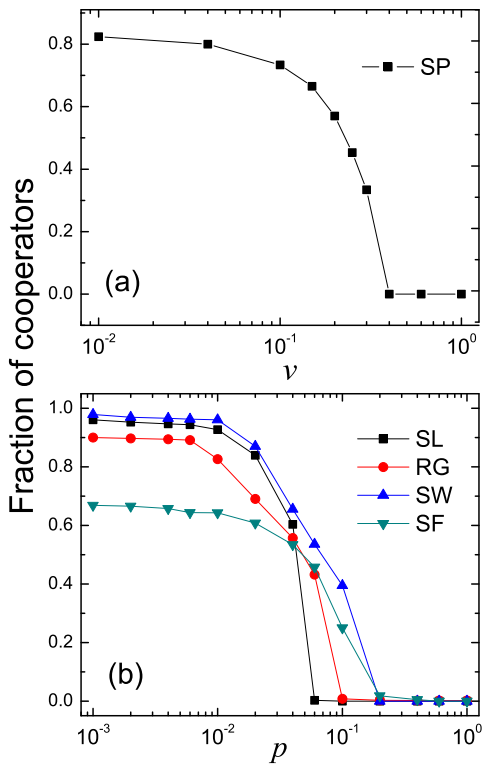}} \caption{(Color online) Fraction of
cooperators as a function of the migration speed $v$ or the
migration probability $p$. (a) Individuals migrate on the square
plane (SP) of linear size $L=20$. The temptation to defect $b=1.35$
and the interaction radius $q=1$. (b) Individuals migrate on various
networks. The temptation to defect $b=1.5$. Average connectivity
$z=4$ for SL and SF, $z=6$ for RG and SW. The population size is
1024. Each data point depicted corresponds to an average over 1,000
simulations; that is, 100 runs for 10 different realizations of the
same class of graph.} \label{fig:migration speed}
\end{center}
\end{figure}

Let us first consider individuals moving on a continuous square
plane (SP) with periodic boundary conditions. Initially, individuals
are randomly located on a square plane and cooperators and defectors
with equal percentage are randomly distributed on the plane. At each
time step, each individual plays the game with individuals falling
in a circle of radius $q$ that centered at his/her current position.
Individuals synchronously update their strategies according to a
best-takes-over reproduction, that is, each individual compares
his/her payoff with his/her neighbors and update his/her strategy by
following the one (including himself/herself) with the greatest
payoff. We have examined that the qualitative results shown below
are robust, regardless of detailed updating rules, such as the
finite population analogue of the replicator dynamics \cite{spatial
5} and Fermi update rule \cite{cluster1,Fermi}. After the strategy
updating process, individuals move to new locations with random
directions of motion in migration speed $v$. The absolute value of
$v$ defines the distance an individual can move in a typical time
step. Simulation results for the fraction of cooperators for
different migration speeds are shown in Fig.~\ref{fig:the fraction
of cooperators}(a). We can see that compared to the static case
($v=0$), cooperation is enhanced in a wide range of temptation to
defect when individuals move slowly ($v=0.04$). On the other hand,
fast moving ($v=1$) leads to complete extinction of cooperators,
analogous to the situation arising in the well-mixed population.

Next, we study individuals migrating on various network models,
including square lattices (SL), random graphs (RG) \cite{RG},
small-world networks (SW) \cite{SW} and scale-free networks (SF)
\cite{SF}. Initially, each node of the network is occupied by an
individual and individuals with two strategies (cooperators or
defectors) are randomly distributed. At each time step, each
individual plays the game with individuals sitting on the same
node and neighboring nodes. Individuals synchronously update their
strategies according to the best-takes-over reproduction and then
each individual jumps to a randomly chosen neighboring node with
probability $p$ (a node can be occupied by more than one
individual). Results are shown in Figs.~\ref{fig:the fraction of
cooperators}(b)-(e). As compared to the static case ($p=0$), low
migration probabilities (e.g. $p=0.001$) promote cooperation in a
wide range of the temptation to defect $b$ (except for large $b$
on square lattices and random graphs, where the fraction of
cooperators is lower than that of $p=0$), similar to the results
on the continuous plane. While for high migration probability
($p=1$), defectors dominate the whole population.

\begin{figure*}
\begin{center}
 \scalebox{0.86}[0.86]{\includegraphics{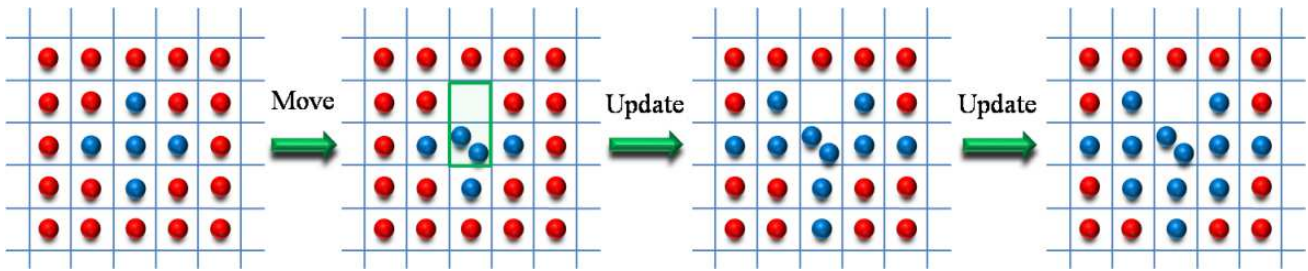}} \caption{Red balls
represent defectors and blue balls denote cooperators. Each
individual plays the game with individuals who are located on the
same lattice and the neighboring lattices. The temptation to defect
$b=1.4$. After a cooperator moves from the cluster boundary into the
core of cooperator cluster, the cooperator cluster expands.
 } \label{fig:sketch map}
\end{center}
\end{figure*}

From Fig.~\ref{fig:the fraction of cooperators}, we can find that by
the comparison with spatial game in the absence of migration, low
migration speeds/probabilities can considerably promote cooperation
whereas high migration speed/probability facilitates defection,
which are qualitatively regardless of underlying structures. We have
also investigated the dependence of fraction of cooperators on the
migration speed $v$ and probability $p$ with fixing the value of
temptation to defect $b$. As exhibited in Fig.~\ref{fig:migration
speed}, as $v$ and $p$ increase, the fraction of cooperation
monotonously decreases. It has been known that in spatial games,
cooperators can survive by forming clusters
~\cite{cluster1,cluster2,cluster3}, in which the benefits of mutual
cooperation can outweigh losses against defectors, thus enable
cooperation to be maintained. Combining Figs.~\ref{fig:the fraction
of cooperators} and \ref{fig:migration speed}, we can find that the
effect of migration on cooperation is twofold. For high migration
speed/probability, cooperation is inhibited since cooperator
clusters can be hardly formed induced by the frequent change of
neighbors. Without the protection of cluster structures, cooperator
can hardly survive. For the low degree of migration, it is not easy
to figure out the influence of migration to cooperation. A heuristic
explanation is that after the construction of cooperator clusters,
small perturbation along the boundary by migration can trigger the
expansion of cooperator clusters and enhance the fraction of
cooperation.

\begin{figure*}
\begin{center}
 \scalebox{0.86}[0.86]{\includegraphics{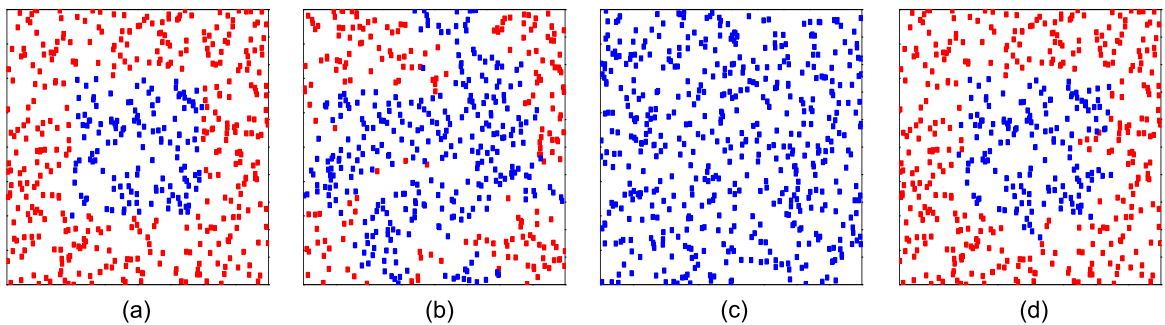}} \caption{(Color online) Snapshots of distributions of cooperators
(blue) and defectors (red) on a square plane of linear size $L=20$.
Initially we set individuals who are located in the middle region of
the plane as cooperators (the cooperative region is square with size
$10\times10$), while defectors are located on other regions. The
population size is 500, the interaction radius $q=1$ and the
temptation to defect $b=1.5$. (a)-(c) Snapshots at different time
steps $t$ for $v=0.1$. (a) $t=0$, (b) $t=263$ and (c) $t=1315$. (d)
The snapshot for $v=0$ when the system reaches equilibrium.
 } \label{time}
\end{center}
\end{figure*}

To intuitively understand the effect of perturbation around the
cooperator cluster on cooperation, we construct a crossed cooperator
cluster (including 5 cooperators) surrounded by defectors on a
square lattice (see Fig. \ref{fig:sketch map}). The temptation to
defect $b=1.4$. If individuals are immobile, the crossed cooperator
cluster is stable and keeps unchanged. In the presence of migration,
situations arising at the cluster boundary can be classified into
four types: (1) a cooperator at the boundary enters the defector
cluster; (2) a defector at the boundary intrudes into the cooperator
cluster; (3) a defector moves away from boundary within its defector
cluster and (4) a cooperator moves away from boundary within its
cooperator cluster. In case (1), the irruptive cooperator transfers
to a defector; In case (2), the irruptive defector changes to a
cooperator; In case (3), nothing happens. Cases (1) to (3) do not
drastically affect fraction of cooperator in the system (not shown
here). However, in case (4), the territory of the cooperator cluster
expands to other regions of the square lattice and the number of
cooperators increases from 5 to 12, as shown in Fig. \ref{fig:sketch
map} \cite{note}. It is thus the rising of case (4) that promotes
the prevalence of cooperation in the population. In general, this
scenario is representative of the strengthening of the cooperator
cluster boundary by multi-cooperators at the same node (the density
of cooperators is augmented along the boundary). A direct result is
that the payoffs of cooperators along the boundary are increased and
defectors nearby the boundary will be assimilated. As a result,
cooperator clusters expand and cooperation is enhanced.

To visually observe how low degree of migration affects the
evolution of cooperator clusters and defector clusters, we initially
set some cooperators in the middle region of a square plane, while
defectors are located on other regions. From Figs.
\ref{time}(a)-(c), one can find that, for low migration speed, the
cooperator cluster gradually expands as time step $t$ increases and
cooperators dominate the whole population in the end. For the static
case in which individuals do not move, the cooperator cluster keeps
almost unchanged [see Fig. \ref{time}(d)].

In summary, we have studied the role of random migration in
cooperation in the framework of spatial prisoner's dilemma
game~\cite{SG} on a variety of spatial structures. Our findings are
that although high degree of migration by disabling the formation of
cooperator clusters results in the extinction of cooperation, low
degree of migration can considerably enhance cooperation by
increasing the cooperator density along the boundary of the
cooperator cluster. Due to the accumulation of cooperators along the
boundary, the benefits of mutual cooperation outweigh losses against
defectors nearby the boundary, this thus not only enables
cooperation within the cluster to be maintained, but also induces
the expansion of the cooperator cluster, in contrast to the static
spatial game. The strengthening at the boundary of cooperator
clusters induced by the small degree of migration plays the key role
in the enhancement of cooperation, regardless of the underlying
structure on which the evolutionary game takes place. Our work may
inspire further effort in exploring the effect of migration behavior
on not only game-based cooperation but also other dynamical
processes, such as epidemic spreading and information routing in
ad-hoc networks.

We thank Hisashi Ohtsuki and Zhi-Xi Wu for helpful comments and
discussions. This work is funded by the National Basic Research
Program of China (973 Program No. 2006CB705500), the National
Natural Science Foundation of China (Grant Nos. 10975126, 10635040),
the Specialized Research Fund for the Doctoral Program of Higher
Education of China (Grant No. 20093402110032).


\begin{references}
\bibitem{note}
Note that for low migration probability $p$, the system has
sufficient time to evolve to another stable state after migration.
\bibitem{SG}
We have found that, the role of migration is the same for the
snowdrift game.
\end{references}
\end{document}